\documentclass[aps,prb,twocolumn,superscriptaddress,longbibliography]{revtex4-2}
\usepackage[utf8]{inputenc}
\DeclareUnicodeCharacter{2009}{\,}
\usepackage{epsfig,amsopn}
\usepackage{graphicx}
\usepackage{physics}
\usepackage{color}
\usepackage{amsmath,amssymb}
\usepackage{enumerate}
\newcommand\bea{\begin{eqnarray}}
\newcommand\eea{\end{eqnarray}}
\newcommand\beq{\begin{equation}}
\usepackage{gensymb}
\usepackage{verbatim}
\newcommand\eeq{\end{equation}}

\def\nn{\nonumber}
\def\f{\frac}
\def\al{\alpha}

\def\si{\sigma}
\def\Do{\partial}
\def\De{\Delta}

\def\ua{\uparrow}
\def\da{\downarrow}

\def\th{\theta}

\def\sq{\sqrt}

\begin{document}
\title{N\'eel vector controlled charge and spin transport in altermagnetic junctions}
\author{Shubham Ghadigaonkar}
\altaffiliation{These authors contributed equally to this work}
%\affiliation{Department of Physics, Tata Institute of Fundamental Research, Homi Bhabha Road, Colaba, Mumbai 400005, India}
\affiliation{School of Physics, University of Hyderabad, Prof. C. R. Rao Road, Gachibowli, Hyderabad-500046, India}

\author{Sachchidanand Das}
\altaffiliation{These authors contributed equally to this work}
\affiliation{School of Physics, University of Hyderabad, Prof. C. R. Rao Road, Gachibowli, Hyderabad-500046, India}

 \author{ Abhiram Soori}
 \email{abhirams@uohyd.ac.in}
 \affiliation{School of Physics, University of Hyderabad, Prof. C. R. Rao Road, Gachibowli, Hyderabad-500046, India}
\begin{abstract}
Altermagnets (AMs) - magnetic materials that have spin-split bandstructure with zero net spin polarization can be classified as weak or strong depending upon the strength of altermagnetic term in the Hamiltonian.  We theoretically investigate electron transport in junctions between the two AMs in strong and weak altermagnetic phases. The charge and spin conductivities are analyzed as functions of angle $\theta$ between the N\'eel vectors of the two AMs. In the strong AM regime, the charge conductivity vanishes as $\theta \to \pi$,  while in the weak AM regime it remains finite. Introducing a normal metal (NM)  between two AMs leads to Fabry–P\'erot-type oscillations in charge conductivity which can be controlled by an applied gate voltage. In the strong regime, transport in AM-NM-AM junctions is dominated by up-spin electrons, whereas both spin channels contribute in the weak regime. These results highlight the potential of AM-based heterostructures for spintronic applications, such as spin filters, and quantum interference–based spintronic devices, where tunable spin-dependent transport and interference effects can be utilized in electronic devices without a need for externally applied magnetic field.
\end{abstract}
%\pacs{}
\maketitle
\section{Introduction}
Altermagnets (AMs), materials having $d$-wave magnetic order have generated tremendous interest among condensed matter physicists in the past couple of years~\cite{smejkal22a,smejkal22b,smejkal22c,fern24,zhou24,yan24,fu2025am,yu25}. Characterized by traits of both ferromagnets and antiferromagnets, their net spin polarization is zero. They are known to carry spin current under a voltage bias~\cite{das2023}.  Junctions of AMs with normal metals, ferromagnets and superconductors have been studied by many groups~\cite{das2023,papaj,sun23,das2024car,fu2025}. The spin that commutes with the Hamiltonian of AM defines the N\'eel vector for the AM.  Several candidate materials such as MnTe, Mn$_5$Si$_3$, KV$_2$Se$_2$O exist for AMs~\cite{Helena21,Jiang2025,Amin2024}. 

Magnetic tunnel junctions (MTJs) are junctions between two ferromagnetic metals separated by a thin insulator, wherein electrons are able to pass through the thin insulating barrier. The relative orientation of the magnetic moments in these layers dictates the possibility of electron tunneling. The parallel  alignment of magnetizations in the ferromagnetic layers exhibits a low electrical resistance in the junction, whereas an antiparallel alignment results in a high resistance. This variation in resistance between the two magnetic configurations gives rise to the tunneling magnetoresistance (TMR) effect, which forms the fundamental basis for the operation of MTJ-based spintronic devices~\cite{Julliere,miyazaki1995, Yuasa2007,Moodera95}. Magnetic tunnel junctions in  altermagnetic RuO$_2$, and  single ferromagnetic electrode have  been shown to result in a high tunnelling magnetoresistance~\cite{Jiang23,Noh25,Samanta24}. Magnetic tunnel junctions between the altermagnets has just recently been studied where the authors claim to achieve tunneling magnetoresistance over 1000\% by just rotating the AM and tuning the altermagnetic strength and Fermi energy~\cite{Sun25,Ezawa25}. 

 The orientation of the N\'eel vector can be tuned using spin–orbit torques or ultrafast optical excitations~\cite{Zhang22,Ono2021}. It influences how spin-polarized currents propagate through the materials.  A recent study on AM/$p-$wave magnet junction by rotating the N\'eel vector of the latter relative to that of the former shows similar response~\cite{das25}. In MnTe, it is found that domains which have different directions for the N\'eel vectors are formed very much like that in ferromagnets~\cite{Amin2024}.
 
 Motivated by these developments, we first study electron transport in  junctions between two altermagnets having different directions of N\'eel vectors modeled by continuum model. We write down boundary conditions that characterize the junction. Then we calculate charge and spin conductivities using the Landauer-B\"uttiker scattering formalism. Also, we sandwich a normal metal (NM) in between the two AMs, having different direction of N\'eel vectors, to study how the inclusion of NM affects the conductivity. 

In Sec.~\ref{sec:calc}, we present the general outline of our calculation. Sections~\ref{sec:AAW} and \ref{sec:ANAW} discuss AM–AM and AM–NM–AM junctions in the weak AM regime. Sections~\ref{sec:AAS} and \ref{sec:ANAS} cover the same junction types in the strong AM regime. Sections~\ref{sec:AAW} and \ref{sec:AAS} contain three parts. In the first part, we give details of calculations, followed by analytical expressions for reflection amplitudes which will be used in calculating the conductivities, and end with a subsection on results and analysis. 
Sections~\ref{sec:ANAW} and  \ref{sec:ANAS} are divided into two parts. The first part contains the detailed calculations, and the second part presents the corresponding results with analysis. In section~\ref{sec:disc}, we compare our work with some closely related works, followed by a discussion of connection to TMR and the effect of disorder and edge roughness. 
Finally, we summarise our findings and provide concluding remarks in Sec.~\ref{sec:sum}.
 
 \section {Outline of calculation}\label{sec:calc}
 A simple model for AMs consists of a Hamiltonian for electrons with spin- and direction- dependent hopping in a two-dimensional square lattice. It breaks time reversal symmetry but is invariant under time reversal times  $\pi/2$-rotation. The Hamiltonian in tight-binding model can be written as sum of two terms: first describing a normal metal and the second describing altermagnetic order. The Hamiltonian can be written as  
 \bea H&=&-2t(\cos k_x a+\cos k_y a)\si_0 \nn \\  && ~~+~~2t_J(\cos{k_x a}-\cos{k_y a})\si_z,\eea
  where $a$ is the lattice spacing, $\si_j$'s are Pauli spin matrices. Here, $t$ is the hopping strength which characterizes a metal, and $t_J$ is spin- and direction- dependent hopping which characterizes altermagnet.  By Taylor expanding such a Hamiltonian in momentum space near the band bottom, its continuum form can be obtained. Depending on the relative strength of the altermagnetic term compared to the normal metal term in the Hamiltonian, the phase of AM can be classified into strong and weak. The strong phase corresponds to $t_J>t\ge 0$ whereas the weak phase corresponds to $0\le t_J<t$. In the weak regime, the Fermi contours for the two spins intersect whereas in the strong phase, they do not. 

  We will consider junctions between two AMs having different N\'eel vectors. The charge  (spin) current density corresponds to the  current density that obeys the continuity equation along with the charge (spin) density defined by $\rho_{\chi}=e\psi^{\dag}\psi$ ($\rho^s_{\chi}=\hbar\psi^{\dag}\si_{\chi}\psi/2$), where $\si_{\chi}=\si_z\cos\chi+\si_x\sin\chi$. The spin density corresponding to the N\'eel vector direction defined by $\chi$ commutes with the Hamiltonian and is different on different sides of the junction.
 
 Charge conservation ensures that the current is continuous across the junction, providing the basis for formulating appropriate boundary conditions at the interface. These boundary conditions can be  mapped to the terms in the Hamiltonian of the underlying lattice Hamiltonian in the vicinity of the junction. Importantly, the resulting boundary conditions differ in the strong and weak altermagnetic regimes. A detailed discussion of the boundary conditions for both cases is presented in the following sections.
 
\section{AM-AM junction in weak phase}\label{sec:AAW}
\subsection{Details of the calculation}
In the weak phase, the location of the band bottoms for both the spins is $\vec k=0$, so, the Hamiltonian for weak phase near band bottom can be written as 
  \bea 
  H_W(\chi)= -(t\si_0-t_J\si_{\chi})a^2\Do^2_x-(t\si_0+t_J\si_{\chi})a^2\Do^2_y .\label{eq:H-weak}
  \eea
Here we consider a junction between two AMs in the weak phase differing in the N\'eel vector directions. To be more precise, in the region left to $x=0$, the Hamiltonian is $H_W(\chi=0)$ and in the region right to $x=0$ the Hamiltonian is $H_W(\chi=\th)$. \\

In the weak AM, dispersion for $\ua$ and $\da$ electrons are given by--
\beq
  E=(t-t_J)k_{x\ua}^2a^2+(t_J+t)k^2_{y\ua}a^2
\eeq
\beq
  E=(t_J+t)k^2_{x\da}a^2+(t-t_J)k_{y\da}^2a^2 \label{eq:disp_wk}
\eeq
The boundary condition that can be derived from the probability current conservation along $\hat x$ is given by 
\bea 
\psi(0^-) &=& c\psi(0^+) \nn \\ 
\big[c(t\si_0-t_J\si_z)a\Do_x\psi + taq_0\psi\big]_{0^-} &=& (t\si_0-t_J\si_{\th})a\Do_x\psi|_{0+} \nn \\
&& \label{eq:bc-weak}\eea
Here, $c$ and $q_0$ characterise the junction. In certain limits, $c$ can be thought of physically as the ratio between the hopping strength at the bond that forms the junction to $t$, and $q_0$ corresponds to the strength of delta-function impurity at the junction~\cite{das2023}.

The scattering eigenfunction for an  $\ua$-electron approaching from the left at energy $E$ and angle of incidence $\al$ can be expressed as $\psi(x)e^{ik_{y\ua}y}$, where 
\bea 
\psi(x) &=& (e^{ik_{x\ua}x}+r_{\ua\ua}e^{-ik_{x\ua}x})\ket{\ua} +r_{\da\ua}e^{-ik_{x\da }x}\ket{\da}, \nn \\ 
&& {\rm for ~~} x<0, \nn \\ 
&=& t_{\ua\ua}e^{ik_{x\ua}x}\ket{\ua_{\th}} +t_{\da\ua}e^{ik_{x\da}x}\ket{\da_{\th}},\nn \\ && {\rm for~~} x>0 .
\eea
Here 
\bea
\ket{\ua}&=&\ket{\ua_{\chi=0}}, \ket{\da}=\ket{\da_{\chi=0}},~
\ket{\ua_{\chi}}=[\cos\f{\chi}{2}, \sin\f{\chi}{2}]^T, \nn \\ ~~\ket{\da_{\chi}}&=&[-\sin\f{\chi}{2}, \cos\f{\chi}{2}]^T , k_{y\ua}=\sqrt{E/(t+t_J)}\sin{\al} \nn \\ k_{x\ua}&=&\sqrt{E/(t-t_J)}\cos{\al}, \nn \\ k_{x\da}&=&\sqrt{\{E/(t+t_J)\}(1-\eta\sin^2\al)} \label{eq:wv_weakup}
\eea and $\eta=(t-t_J)/(t+t_J)$. The scattering coefficients $r_{\si'\si}$ and $t_{\si'\si}$ can be found using the boundary conditions in Eq.~\eqref{eq:bc-weak}, where $\si=\ua$ and $\si'=\ua {\rm or} \da$.

The charge- and spin-  current densities in the system due to this wavefunction are given by 
\bea 
J^c_{\ua}(\al) &=& \f{2e}{\hbar}\big((t-t_J)k_{x\ua}(1-|r_{\ua\ua}|^2)-(t+t_J)k_{x\da}|r_{\da\ua}|^2\big) ,\nn \\ 
J^{s-}_{\ua}(\al) &=& (t-t_J)k_{x\ua}(1-|r_{\ua\ua}|^2)+(t+t_J)k_{x\da}|r_{\da\ua}|^2 ,\nn \\ 
J^{s+}_{\ua}(\al) &=& (t-t_J)k_{x\ua}|t_{\ua\ua}|^2-(t+t_J)k_{x\da}|t_{\da\ua}|^2 . \label{eq:Jup-weak}
\eea
Note that while the charge current is same in the two AMs, the spin current need not be the same, since none of the Pauli spin matrices commute with Hamiltonians on both sides. $J^{s+}$ ($J^{s-}$) is the spin current density on the right (left) side of the junction. While the spin current density on the left corresponds to $\si_z$, the one on the right corresponds to $\si_{\th}$.

The scattering eigenfunction corresponding to a $\da$-electron incident at an angle of incidence $\al$ at energy $E$ has the form $\psi(x)e^{ik_{y\da}y}$ where, 
\bea 
\psi(x) &=& (e^{ik_{x\da}x}+r_{\da\da}e^{-ik_{x\da}x})\ket{\da} +r_{\ua\da} e^{-ik_{x\ua}x}\ket{\ua}, \nn \\ 
&& {\rm for ~~}x<0 \nn \\ 
&=& t_{\ua\da}e^{ik_{x\ua}x}\ket{\ua_{\th}} +t_{\da\da}e^{ik_{x\da}x}\ket{\da_{\th}} \nn \\ 
&& {\rm for~~} x>0 .
\eea
 Here, 
 \bea
 k_{y\da}&=&\sqrt{E/(t-t_J)}\sin\al, k_{x\da}=\sqrt{E/  (t+t_J)}\cos\al,\nn \\ k_{x\ua}&=&\sqrt{E/(t-t_J)}\sqrt{1-(\sin^2\al)/\eta} \label{eq:wv_weakdn}
 \eea
 When $\sin^2\alpha/\eta>1$, $k_{x\ua}$ is imaginary.  It is chosen in such a way that the wavefunction for $\ua$-electron decays away from the junction.  Using the boundary conditions shown in Eq.~\eqref{eq:bc-weak}, the scattering coefficients $r_{\si'\si}$ and $t_{\si'\si}$ can be calculated, where $\si=\da$ and $\si'=\da {\rm or} \ua$.
 
 This wavefunction results in the following charge and spin current densities
 \bea 
 J^c_{\da}(\al) &=&  \f{2e}{\hbar}\big((t+t_J)k_{x\da}(1-|r_{\da\da}|^2)\nn \\ && -(t-t_J){\rm Re}[k_{x\ua}] |r_{\ua\da}|^2 \big) ,\nn \\ 
J^{s-}_{\da}(\al) &=&  -(t+t_J)k_{x\da}(1-|r_{\da\da}|^2)\nn \\ &&-(t-t_J){\rm Re}[k_{x\ua}] |r_{\ua\da}|^2  ,\nn \\ 
J^{s+}_{\da}(\al) &=& (t-t_J){\rm Re}[k_{x\ua}]|t_{\ua\da}|^2\nn \\ &&-(t+t_J)k_{x\da}|t_{\da\da}|^2 \label{eq:Jdn-weak}
 \eea
 
 The differential charge and the spin conductivities are given by 
 \bea
 G &=& \f{e}{8\pi^2\sqrt{t^2-t_J^2}}\int_{-\pi/2}^{\pi/2}d\al [J^c_{\ua}(\al)+J^c_{\da}(\al)], \nn \\ 
 G^{s\pm} &=& \f{e}{8\pi^2\sqrt{t^2-t_J^2}}\int_{-\pi/2}^{\pi/2}d\al [J^{s\pm}_{\ua}(\al)+J^{s\pm}_{\da}(\al)] \nn\\ &&\label{eq:cond_wk}
 \eea
 \subsection{Analytical expressions in the limit $\th=0$}
 In the limit $\th=0$,  it is straightforward to show that  $r_{\ua\da}=r_{\da\ua}=0$ and 
\bea
r_{\ua\ua} &=& \f{tq_0 +ik_{x\ua}(t-t_J)(1-1/c^2)}{ik_{x\ua}(t-t_J)(1+1/c^2)-tq_0} \nn \\
r_{\da\da} &=& \f{tq_0 +ik_{x\da}(t+t_J)(1-1/c^2)}{ik_{x\da}(t+t_J)(1+1/c^2)-tq_0}
\eea
 
 \subsection{Results}
\begin{figure}[htb]
\includegraphics[width=8cm]{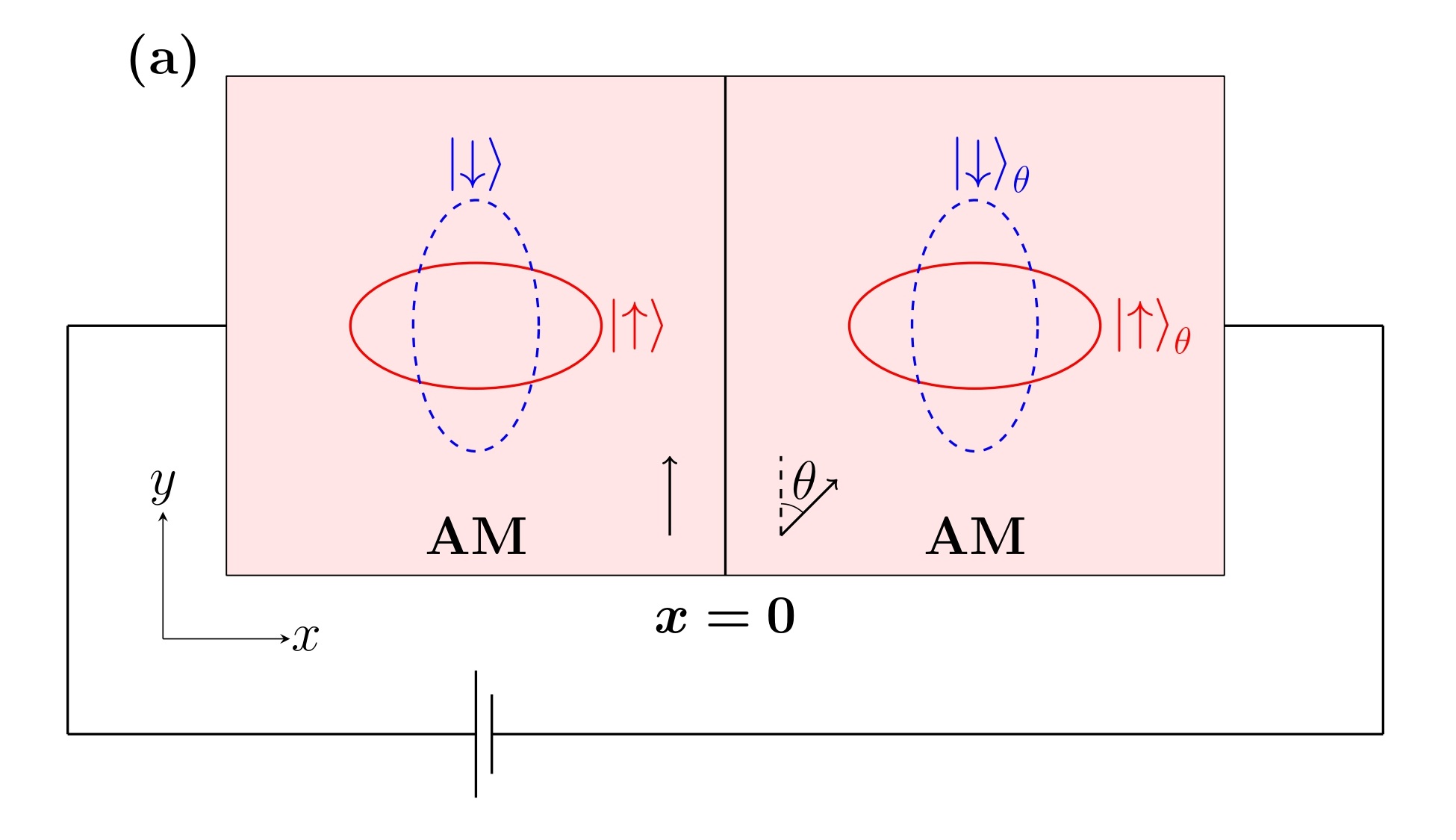}
\includegraphics[width=4cm]{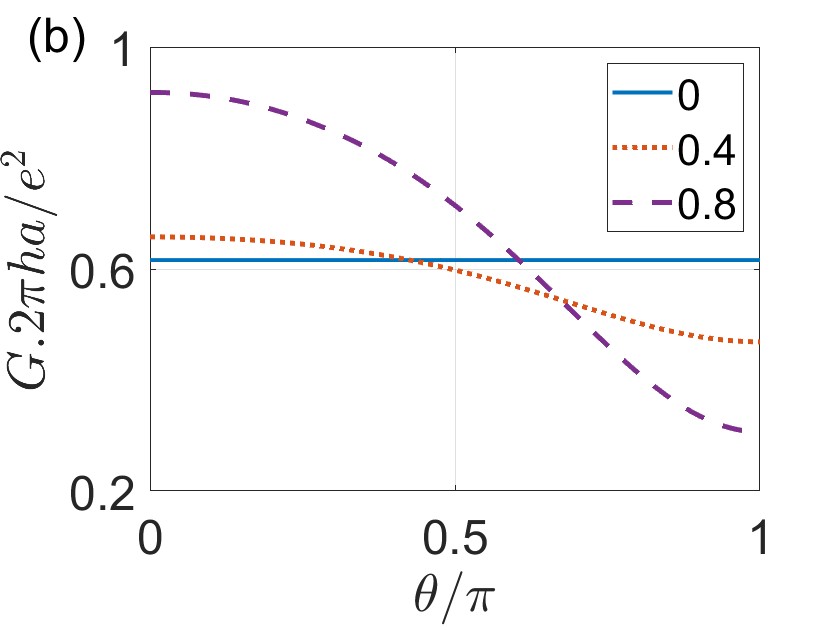}
\includegraphics[width=4.5cm]{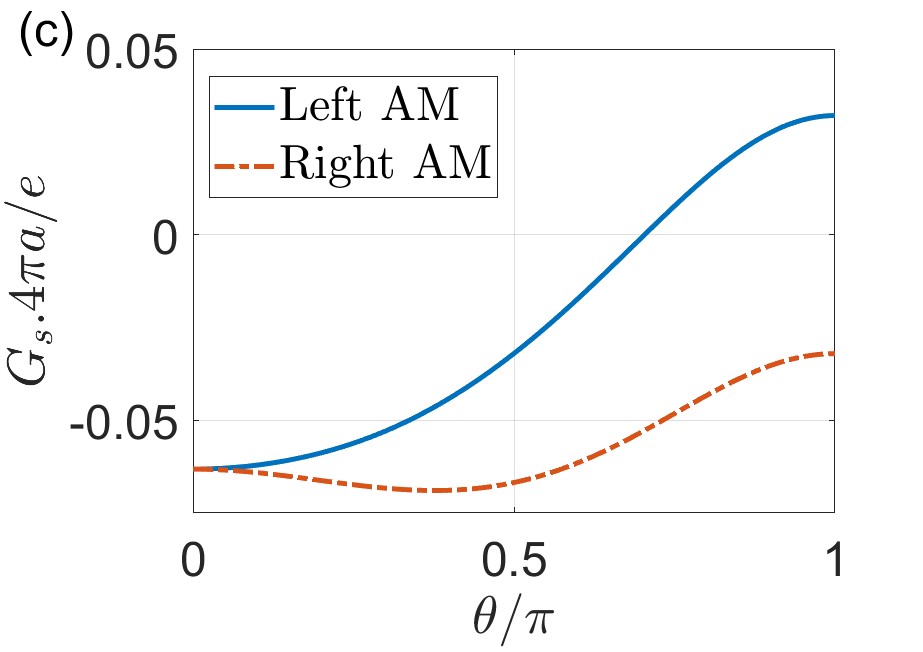}
\caption{(a) Schematic of the junction with the Fermi surfaces on each region. The N\'eel vectors on either sides of the junction differ by an angle $\th$. (b) Differential charge conductivity versus $\th$ for different values of $t_J/t$ indicated in the legend. Other parameters: $q_0=1/a$, $c=1$, $E=t$ (c) Spin conductivity versus $\th$ in the left and right AM for $q_0=0$, $c=1.2$, $t_J=0.2$ and $E=t$.} \label{fig:weak}
\end{figure}

 In Fig.~\ref{fig:weak}(b,c), we plot the conductivities versus the angle between the N\'eel vectors of the two AMs. In Fig.~\ref{fig:weak}(b), the charge conductivity is shown wavevector the angle $\th$ for different ratios of $t_J/t$, with parameters $q_0=1/a$, $c=1$, and $E=t$. For larger values of the altermagnetic strength $t_J$, the conductivity exhibits significant  variation. We see [Fig.~\ref{fig:weak}(b)] that for $\th=0$, the charge conductivity is maximum and then decreases monotonically as $\th$ increases in the range $[0,\pi]$. As $\th$ deviates away from $0$, the orientations of the spins corresponding to the same values of $\vec k$ on either sides of the junction  is different leading to reduced conductivity.  Interestingly, the conductivity at $\th=0$ increases with increasing value of $t_J$. This feature can be understood by taking the limit where all the reflection coefficients are zero and using Eq.~\eqref{eq:cond_wk}. 
 As $t_J$ decreases, this variation becomes progressively weaker, and at $t_J=0$, the conductivity remains constant across all angles. This behaviour corresponds to the complete absence of altermagnetic effects, rendering the system equivalent to a normal metal.

Fig.~\ref{fig:weak}(c) illustrates the variation of spin conductivities in the left and right AM regions wavevector $\th$, for parameters $q_0=0$, $c=1.2$, $t_J=0.2$, and $E=t$. The spin conductivity, defined as the difference between the up-spin and down-spin conductivities, has the same value in both regions at $\th=0$. It gradually increases with $\th$ in both the regions. In the left region, the spin current remains mostly negative, while in the right region it stays entirely negative across all $\th$, indicating that down-spin contributions dominate over up-spins.  This is because, the down-spin Fermi surface is elongated along $y$ and  there are more  states that are closer to normal incidence than those for the up-spin incidence wherein the Fermi surface is elongated along $x$.  

At $\th=\pi$, the spin conductivities in the left and right regions shift symmetrically above and below zero.
This can be understood as follows. The left and right AM regions are structurally identical, differing only in the orientation of their magnetization by an angle $\th$. At $\th = 0$, both regions have their magnetization aligned in the same direction, leading to identical up-spin and down-spin transport channels. As a result, the difference between these channels—and hence the spin conductivity—is the same on both sides. However, at $\th = \pi$, the magnetization in the right AM is completely reversed relative to the left AM, effectively corresponding to a spin inversion. In this case, the up-spins in the left region correspond to the down-spins in the right region and vice versa, producing spin conductivities of equal magnitude but opposite sign.

%$$$$$$$$$$$$$$$$$$$$$$$$$$$$$$$$$$$$$$$$$$$$$$$$$$$$
 \section{AM-NM-AM junction in weak phase}\label{sec:ANAW}
 \subsection{Details of the calculation}

 Now a normal metal (NM) of length $L$ is sandwiched between the two AMs having different N\'eel vectors differing by $\chi$.  The region to the left of $x<0$, the Hamiltonian is $H_W(\chi=0)$ and to the right of  $x>L$ is $H_W(\chi=\theta)$. In the region $0<x<L$, the Hamiltonian is given by $H_W(\chi=0,t_J=0)$.

Dispersion for $\ua$- and $\da$ spin electrons in the AM is given in Eq. \ref{eq:disp_wk}. But the dispersion of the normal metal is given by 
\beq
  E=t_0a^2(q_x^2+q_y^2)-\mu
\eeq
\begin{comment}
Current probability densities at the three different regions are given below-
\beq
  J_{L,W}=\f{2}{\hbar}\Big[ {\rm Im\big(\psi_L^{\dagger}(t_0\sigma_0-t_J\sigma_z})\Do_x\psi_L\big)\Big] \nn  
\eeq
\beq
J_{N}=\f{2}{\hbar}\Big[ {\rm Im \big(\psi^{\dagger}_{NM}}t_0\sigma_0 \Do_x\psi_{NM}\big)\Big] \nn 
\eeq
\beq
J_{R,W}=\f{2}{\hbar}\Big[ {\rm Im\big(\psi_R^{\dagger}(t_0\sigma_0-t_J\sigma_{\theta}})\Do_x\psi_R\big)\Big]  
\eeq
\end{comment}
Conservation of probability current density  at the junction of left  AM/NM junction at $x=0$ and right AM/NM junction at $x=L$ gives  boundary conditions
\bea 
\psi(0^-) &=& c\psi(0^+) \nn \\ 
c\big[(t\si_0-t_J\si_z)a\Do_x\psi\big]_{0^-} &=& \big[t_0\si_0(a\Do_x-aq_0)\psi\big]_{0^+} \nn \\
\psi(L^-) &=& c\psi(L^+) \nn \\ 
c\big[t_0(\si_0a\Do_x+aq_0)\psi\big]_{L^-} &=& \big[(t\si_0-t_J\si_{\theta})a\Do_x\psi\big]_{L^+} \nn \\
&& \label{eq:bc-weak_NM}\eea
The scattering eigenfunction corresponding to a $\ua$-electron with energy $E$, incident at an angle $\al$, takes the form $\psi(x)e^{ik_{y\ua}y}$
\bea 
\psi(x) &=& (e^{ik_{x\ua}x}+r_{\ua\ua}e^{-ik_{x\ua}x})\ket{\ua} +r_{\da\ua}e^{-ik_{x\da }x}\ket{\da}, \nn \\ 
&& {\rm for ~~} x<0, \nn \\ 
&=& \Big(A_R e^{iq_xx} + A_L e^{-iq_xx}\Big)\ket{\ua}+\Big(B_R e^{iq_xx} + \nn \\ &&  B_L e^{-iq_xx}\Big)\ket{\da} ~~~~ {\rm for~~} 0<x<L \nn \\
&=& t_{\ua\ua}e^{ik_{x\ua}x}\ket{\ua_{\th}} +t_{\da\ua}e^{ik_{x\da}x}\ket{\da_{\th}},\nn \\ && {\rm for~~} x>L .
\eea
where the wavevectors in the two AMs are given by the Eq.\eqref{eq:wv_weakup} and $q_xa=\sq{(E+\mu)/t_0-k_{y\ua}^2}$. The scattering coefficients $A_R,~B_R,~A_L,B_L,~r_{\si'\si}$ and $t_{\si'\si}$ can be found using the boundary conditions in Eq.~\eqref{eq:bc-weak_NM}.

The charge- and spin-  current densities on two sides of the junction due to this wavefunction are given by Eq. \eqref{eq:Jup-weak} where $J^{s-}_{\ua}(\al)$ and $J^{s+}_{\ua}(\al)$ are the spin current for the left and right AM respectively.

Now when a down-spin electron is incident from the left AM, the scattering eigenfunction of it with energy $E$, incident at an angle $\al$, takes the form $\psi(x)e^{ik_{y\da}y}$
\bea 
\psi(x) &=& (e^{ik_{x\da}x}+r_{\da\da}e^{-ik_{x\da}x})\ket{\da} +r_{\ua\da}e^{-ik_{x\ua }x}\ket{\ua}, \nn \\ 
&& {\rm for ~~} x<0, \nn \\ 
&=& \Big(A_R e^{iq_xx} + A_L e^{-iq_xx}\Big)\ket{\ua}+\Big(B_R e^{iq_xx} +  \nn \\ &&  B_L e^{-iq_xx}\Big)\ket{\da}~~~~{\rm for~~} 0<x<L \nn \\
&=& t_{\ua\da}e^{ik_{x\ua}x}\ket{\ua_{\th}} +t_{\da\da}e^{ik_{x\da}x}\ket{\da_{\th}},\nn \\ && {\rm for~~} x>L .
\eea
here the wavevectors in the two AMs are given by the Eq.\eqref{eq:wv_weakdn} and $q_xa=\sq{(E+\mu)/t_0-k_{y\da}^2}$. The scattering coefficients $A_R,~B_R,~A_L,B_L,~r_{\si'\si}$ and $t_{\si'\si}$ can be found using the boundary conditions in Eq.~\eqref{eq:bc-weak_NM}. \\
The charge- and spin- current densities on two sides of the junction due to this wavefunction are given by Eq. \eqref{eq:Jdn-weak}, where $J^{s-}_{\da}(\al)$ and $J^{s+}_{\da}(\al)$ are the spin current for the left and right AM respectively.\\
The differential charge and spin conductivities in this system are given  by Eq.\eqref{eq:cond_wk}.
\subsection{Results}
\begin{figure}[htb]
\includegraphics[width=9cm]{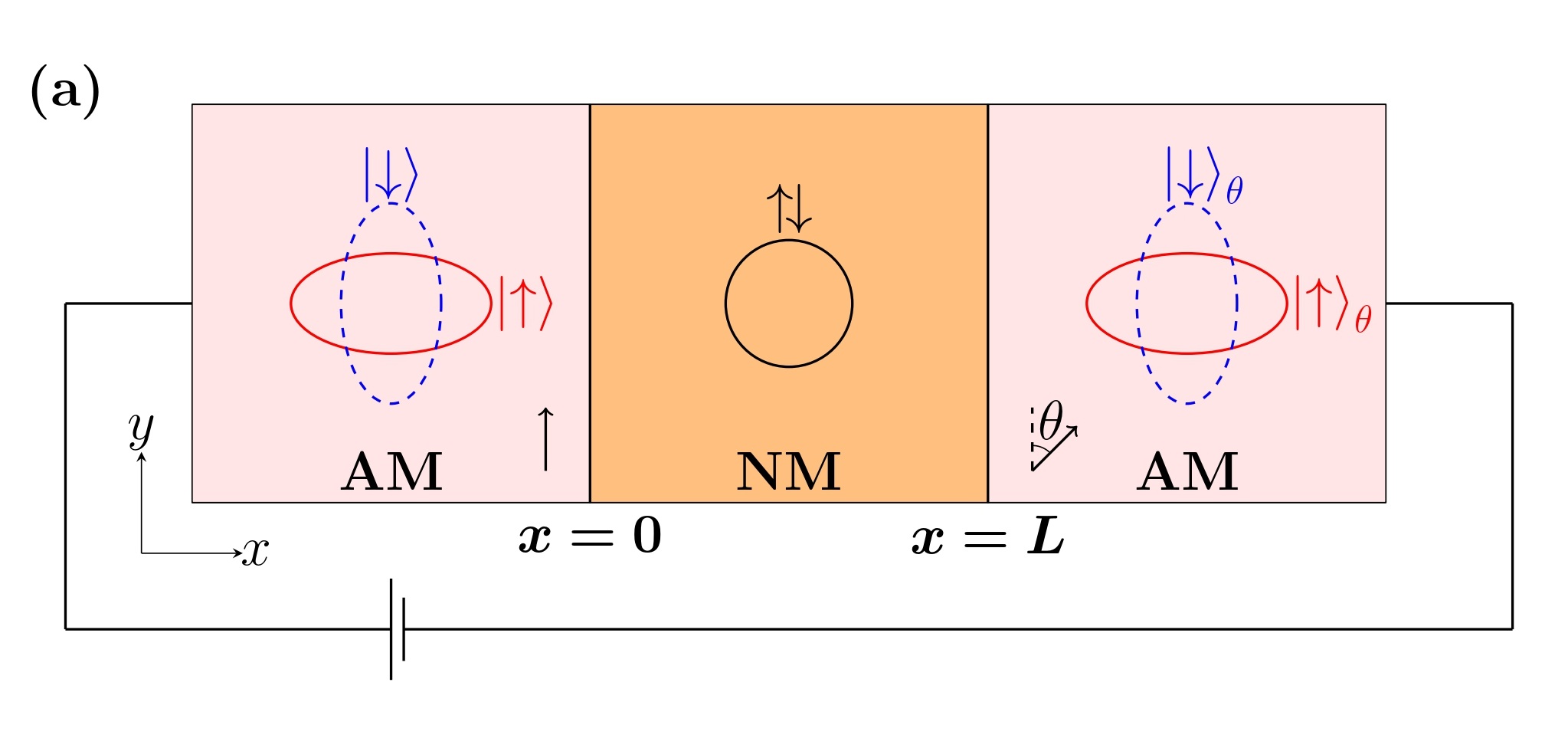}
\includegraphics[width=4.2cm,height=3.5cm]{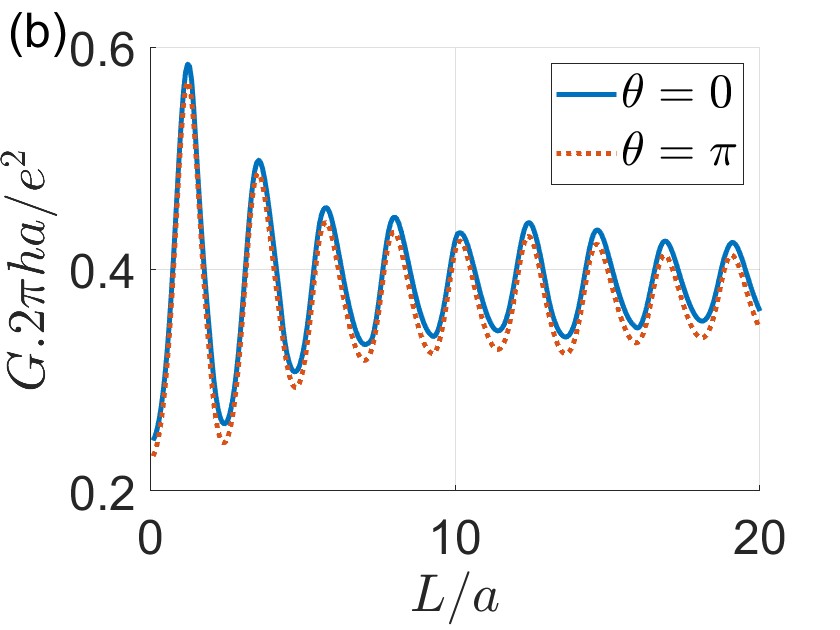}
\includegraphics[width=4.2cm,height=3.5cm]{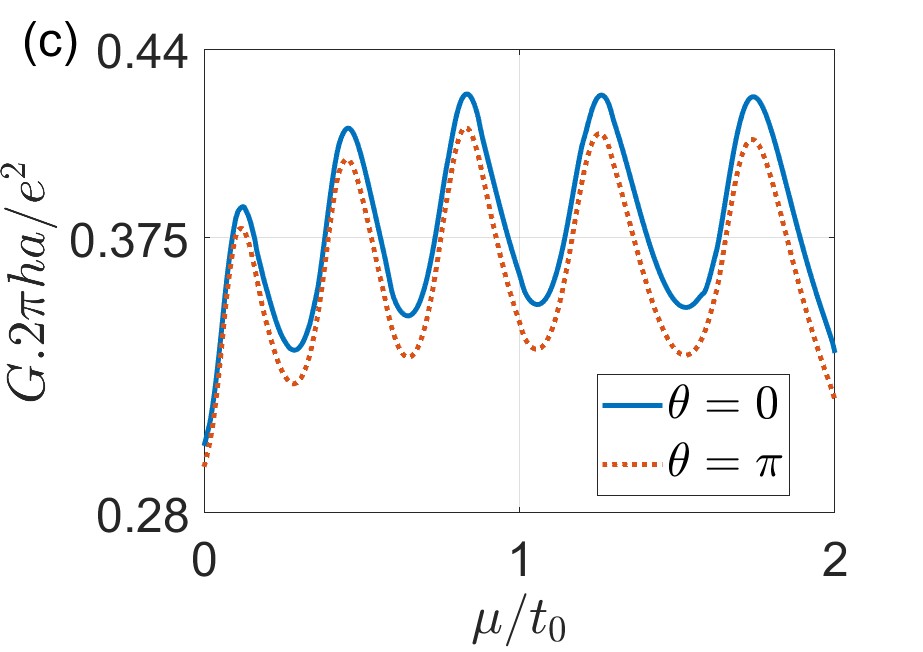}
\caption {(a) Schematic of the system.  Fermi surfaces in each region are indicated by curves. The N\'eel vectors on the left AM and right AM differ by an angle $\th$.  Differential conductivity (b) versus $L$ in the units of $a$ keeping $\mu=t_0$, (c) versus $\mu$ in the units of $t_0$ keeping $L=20$ for two different values of $\th$ i.e $\th=0~{\rm and}~~\pi$  indicated in the legend. Other parameters: $q_0=1/a$, $c=1$, $t_J=0.2t$, $E=t$ are same for (b) and (c)}\label{fig:weak_NM}
\end{figure}
Figure~\ref{fig:weak_NM}(b) shows variation of total conductivity with respect to the Length $(L)$ of the NM at $\theta=0$ (blue solid line) and at $\theta=\pi$ (red dotted line). The conductivity shows an oscillatory dependence on the length  $L$, where the oscillation amplitude initially is large at lower $L$ and then decreases, eventually stabilizing with an almost constant amplitude for larger $L$. This behaviour arises because, for $k_y$ away from $0$, the down spin electrons from AM do not have plane wave states on the NM [see Fig.~\ref{fig:weak}(a)], making the states evanescent. 
For small $L$, electrons are transmitted through evanescent waves and contribute to the total conductivity. However, as $L$ increases, the contribution from spin-down electrons with large $k_y$ vanishes. 

 Such oscillations originate from quantum‐interference effects: multiple reflections within the finite NM segment generate constructive or destructive interference determined by the accumulated phase, known as Fabry-P\'erot interference(FPI)~\cite{soori12,liang2001} . As $L$ increases, the phase acquired by the electron wavefunction varies, leading to alternating constructive and destructive interference.  FPI condition given by $\De L=\pi/q$, where $\De L$ is the interval between successive peaks and $q$ is the wave number of the interfering mode in the NM. This condition is obtained by considering the normal incident electrons which are the dominant contributions to conductivity. $\De L$ calculated by the above condition is 2.221a, in  agreement with the numerically obtained value of 2.212a  obtained in the results in Fig.~\ref{fig:weak_NM}(b).

While the oscillatory dependence on NM length is present for both spin configurations due to quantum interference, the relative amplitude   difference in overall magnitude between the two spin orientations originates from the spin-dependent tunneling at the AM/NM interfaces. Fermi surface shows that transverse momentum matching  is higher for up-spin electrons as compared to the $\da$ electrons at AM/NM interface as all the $k_y$ for up-spin electrons matches with the $k_y$ for NM, but for down-spin it is not the case. So, for $\theta=0$, the spin polarizations of the left and right AM are collinear, which maximizes the effective overlap (both up and down) between the transmitted and incident spin states across the junction. For this case most of the current is carried by $\ua$ electrons leading to higher conductivity. In contrast, for $\theta=\pi$, the spin orientations of the two AMs are antiparallel. So most of the current is carried by the down-spin electrons. Since there are less number of $k_y$ states for down-spin electrons to carry current, we observe a slightly lower conductivity. 

Figure~\ref{fig:weak_NM}(c) presents the variation of the total conductivity wavevector the chemical potential ($\mu$) of the normal metal at two different values of $\th$. Similar to the case of varying NM length, the conductivity for both $\theta=0$ and $\theta=\pi$ exhibits oscillations due to FPI. However, the oscillation amplitude here remains nearly constant after an initial increase.  The initial increase in conductivity is due to the increase in the size of the Fermi surface of the NM, which accommodates more electrons from AM. 

The FPI condition is $\De q=\pi/L$ at fixed length $L$, where $\De q$ is the interval between the successive peaks at $\mu_1$ and $\mu_2$. $\De q$ calculated by this condition is 0.157/a which closely matches with 0.148/a that is obtained in the results in Fig.~\ref{fig:weak_NM}(c) by $\De q=\sqrt{(E+\mu_2)/{t_0}}-\sq{(E+\mu_1)/t_0}$.

 \section{AM-AM junction in strong phase}\label{sec:AAS}
  \subsection{Details of the calculation}

 In the strong phase, electrons of the two spins have different band bottoms. For $\ua$-the band bottom lies at $k_x=\pi/a, k_y=0$ whereas for $\da$- band bottom lies at $k_x=0, k_y=\pi/a$. So the Hamiltonian near the band bottom for the strong phase can be written as
\bea 
H_S(\chi)&=& -\big[(t_J-t)\big(\Do_x-i\f{\pi}{a}\big)^2+(t_J+t)\Do_y^2\big]a^2 \ket{\ua_{\chi}}\bra{\ua_{\chi}} \nn \\ 
&& -\big[(t_J+t)\Do^2_x+(t_J-t)\big(\Do_y \pm i\f{\pi}{a}\big)^2\big]a^2 \ket{\da_{\chi}}\bra{\da_{\chi}}, \nn \\ 
&& \label{eq:H-str}
\eea

To the left of $x=0$, the Hamiltonian is $H_S(\chi=0)$ and to the right of $x=0$ the Hamiltonian is $H_S(\chi=\th)$.\\
Dispersion of altermagnet in the strong phase for up-spin and down-spin electrons are given by--
\beq
  E=(t_J-t)(k_{x\ua}a - \pi)^2+(t_J+t)k^2_{y\ua}a^2
\eeq
\beq
  E=(t_J+t)k^2_{x\da}a^2+(t_J-t)(k_{y\da}a \mp \pi)^2
\eeq
Probability current density for $x<0$ and $x>0$ respectively is given by-
 \bea
   J^-_{Strong}=\frac{2}{\hbar} {\rm Im}\Bigg[\psi^{\dagger}\Big\{(t_J\sigma_0-t\sigma_z)(\Do_x-i\f{\pi}{a}\sigma_{\ua})\psi\Big\}\Bigg]  \\
   J^+_{Strong}=\frac{2}{\hbar} {\rm Im}\Bigg[\psi^{\dagger}\Big\{(t_J\sigma_0-t\sigma_{\theta})(\Do_x-i\f{\pi}{a}\sigma_{\ua\theta})\psi\Big\}\Bigg]
 \eea
 where $\sigma_{\ua}=(\sigma_0+\sigma_z)/2$, $\sigma_{\ua \theta}=(\sigma_0+\sigma_{\theta})/2$.
The conservation of probability current across the junction provides the necessary boundary conditions and is given below -
\bea 
\psi(0^-) &=& c\psi(0^+) \nn \\ 
 c\Big[\big(t_J\sigma_0&-&t\sigma_z)(\Do_x\psi-i\f{\pi}{a}\sigma_{\ua}\psi\big)+q_0\psi\Big]_{0^-}  \nn \\ &=&\Big[\big(t_J\sigma_0-t\sigma_{\theta})(\Do_x\psi  -i\f{\pi}{a}\sigma_{\ua \theta}\psi\big)\Big]_{0^+}\nn \\
&& \label{eq:bc-strong}\eea

When an $\ua$-spin electron with energy $E$ is incident from the left AM making an angle $\al$ with $x$-axis at the junction, the scattering wavefunction associated with the electron has the form $\psi=\psi(x)~e^{ik_{y\ua}y}$, where 
\bea 
\psi(x) &=& (e^{ik_{x\ua}x}+r_{\ua\ua}e^{i(2\pi/a-k_{x\ua})x})\ket{\ua} +r_{\da\ua}e^{-ik_{x\da }x}\ket{\da}, \nn \\ 
&& {\rm for ~~} x<0, \nn \\ 
&=& t_{\ua\ua}e^{ik_{x\ua}x}\ket{\ua_{\th}} +t_{\da\ua}e^{ik_{x\da}x}\ket{\da_{\th}},\nn \\ && {\rm for~~} x>0 .
\eea
where 
\bea
  k_{x\ua}a&=&\pi+\sqrt\frac{E}{t_J-t} \cos\al,~
  k_{y\ua}a=\sqrt\frac{E}{t_J+t}\sin\al~~~\nn {\rm and} \\
  k_{x\da}a&=&\sqrt{\f{E}{t+t_J}-\eta\Big(\sq{\f{E}{t+t_J}}\sin{\al-\pi~sgn(\al)\Big)^2}}~~\label{eq:wv_strng}
\eea
For the $\ua$-spin incident electrons, $k_{x\da}a$ becomes imaginary, that means  there is no $k_y$ for the $\da$-spin reflected and transmitted electrons which matches with the $k_y$ of incident electrons.
The charge and spin current densities corresponding to this wavefunction are given by - 
 \bea 
 J^c_{\ua}(\al) &=&  \f{2e}{\hbar}\Big[(t_J-t)(k_{x\ua}-\pi/a)(1-|r_{\ua\ua}|^2)\nn \\ && -(t_J+t)Re(k_{x\da}) |r_{\da\ua}|^2 \Big] ,\nn \\ 
J^{s-}_{\ua}(\al) &=&  (t_J-t)(k_{x\ua}-\pi/a)(1-|r_{\ua\ua}|^2)\nn \\ && +(t_J+t)Re(k_{x\da})|r_{\da\ua}|^2 ,\nn \\ 
J^{s+}_{\ua}(\al) &=& (t_J-t)(k_{x\ua}-\pi/a)|t_{\ua\ua}|^2\nn \\ && -(t_J+t)Re(k_{x\da})|t_{\da\ua}|^2  \label{eq:Jup-strng}
 \eea
 
Similarly scattering eigenfunction for a $\da$ electron, having energy $E$, being incident at an angle $\al$ with $x$-axis takes the form $\psi=\psi(x)~e^{ik_{y\da}y}$,
 \bea 
\psi(x) &=& (e^{ik_{x\da}x}+r_{\da\da}e^{-ik_{x\da}x})\ket{\da} +r_{\ua\da} e^{i(2\pi/a-k_{x\ua})x}\ket{\ua}, \nn \\ 
&& {\rm for ~~}x<0 \nn \\ 
&=& t_{\ua\da}e^{ik_{x\ua}x}\ket{\ua_{\th}} +t_{\da\da}e^{ik_{x\da}x}\ket{\da_{\th}} \nn \\ 
&& {\rm for~~} x>0 .
\eea
where 
\bea
k_{x\da}a&=&\sqrt\frac{E}{t_J+t} \cos\al,~k_{x\ua}a=\pi + \sqrt{\f{E}{t_J-t}+\f{1}{\eta}k_{y\da}^2a^2} \nn \\
k_{y\da}a&=&-sgn(\al)~\f{\pi}{a}+\sqrt\frac{E}{t_J-t}\sin\al~~~~\label{eq:wv_weak}
\eea
Similar to the above case, here $k_{x\ua}$ becomes imaginary for $\da-$spin incidence exhibiting the same physical interpretation but with opposite spin.\\
Spin current densities across both the sides of the junction is given below. Since charge current is conserved on both regions across the junction,  only the charge current on the left AM is shown below.
 \bea 
 J^c_{\da}(\al) &=&  \f{2e}{\hbar}\Big[(t_J-t){\rm~Re}(\pi/a-k_{x\ua})    |r_{\ua\da}|^2)\nn \\ && +(t_J+t)k_{x\da} (1-|r_{\da\da}|^2) \Big] ,\nn \\ 
J^{s-}_{\da}(\al) &=&  (t_J-t){\rm~Re}(\pi/a-k_{x\ua})|r_{\ua\da}|^2\nn \\ && -(t_J+t)k_{x\da} (1-|r_{\da\da}|^2)  ,\nn \\ 
J^{s+}_{\da}(\al) &=& (t_J-t){\rm Re}(k_{x\ua}-\pi/a)|t_{\ua\da}|^2\nn \\ && -(t_J+t)k_{x\da}|t_{\da\da}|^2  \label{eq:Jdn-strng}
 \eea

  The differential charge and the spin conductivities are given by Eq.~\eqref{eq:cond_wk}
\subsection{Analytical expressions in the limit $\th=0$}
In the limit $\th=0$, the reflection amplitudes $r_{\ua\da}=r_{\da\ua}=0$ and 
\bea 
r_{\ua\ua} &=& \f{[q_0t+i(t_J-t)(k_{x\ua}-\pi/a)(1-1/c^2)]}{[-q_0t+i(t_J-t)(k_{x\ua}-\pi/a)(1+1/c^2)]} \nn \\ 
r_{\da\da} &=& \f{[q_0t+ik_{x\da}(t_J+t) (1-1/c^2)]}{[-q_0t+ik_{x\da}(t_J+t)(1+1/c^2)]} 
\eea

\subsection{Results}
\begin{figure}[htb]
\includegraphics[width=8cm]{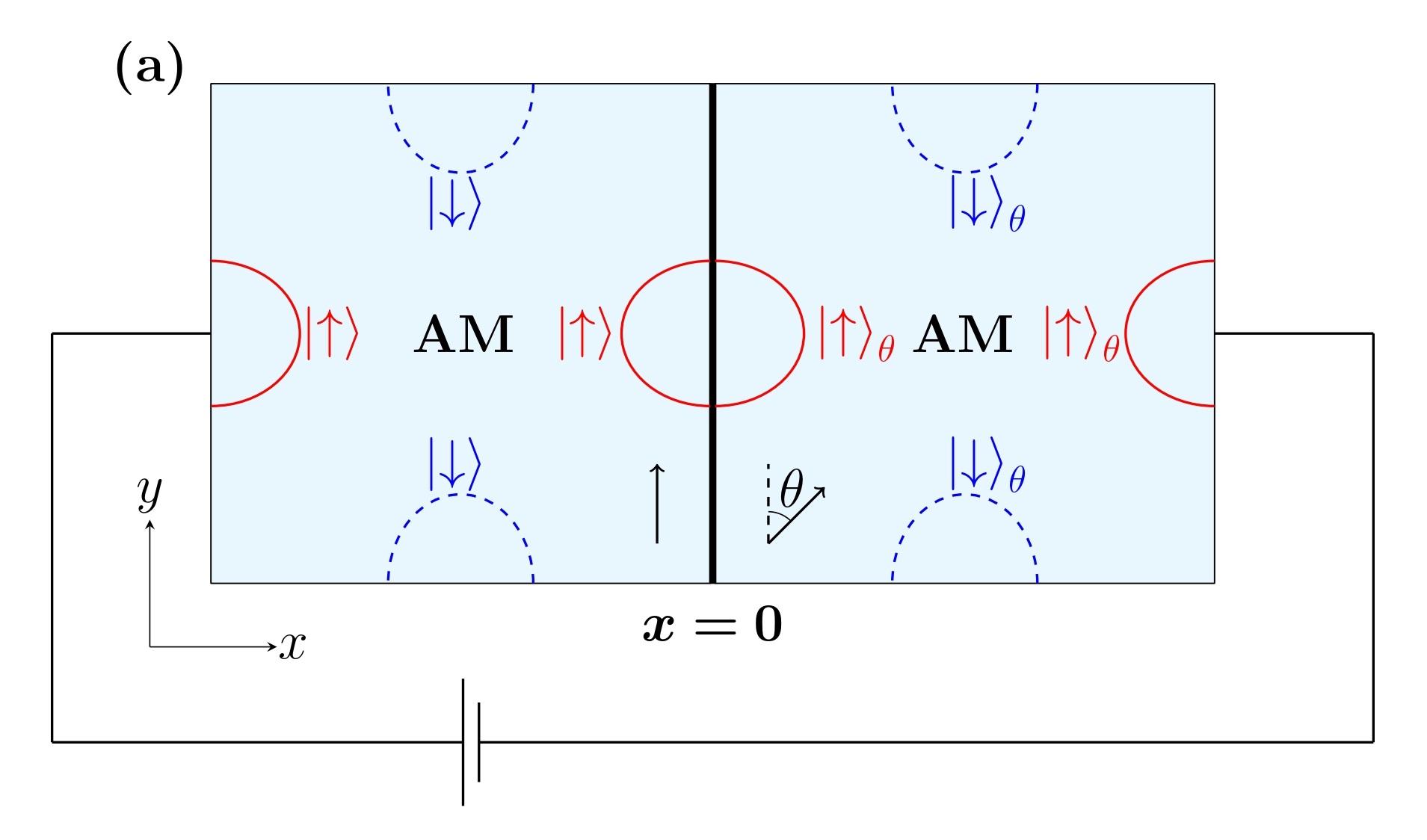}
\includegraphics[width=4.2cm,height=3.5cm]{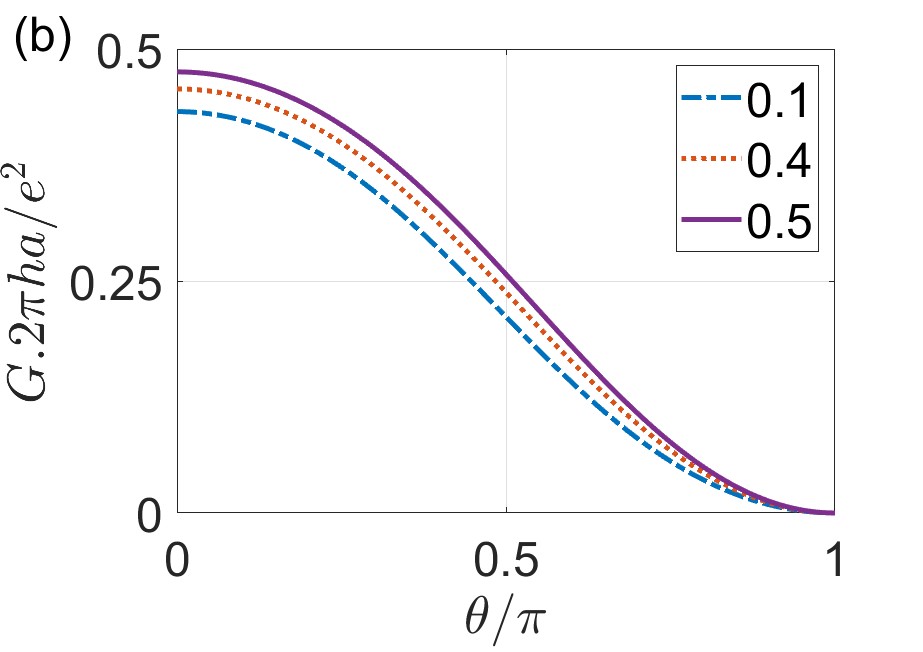}
\includegraphics[width=4.2cm,height=3.5cm]{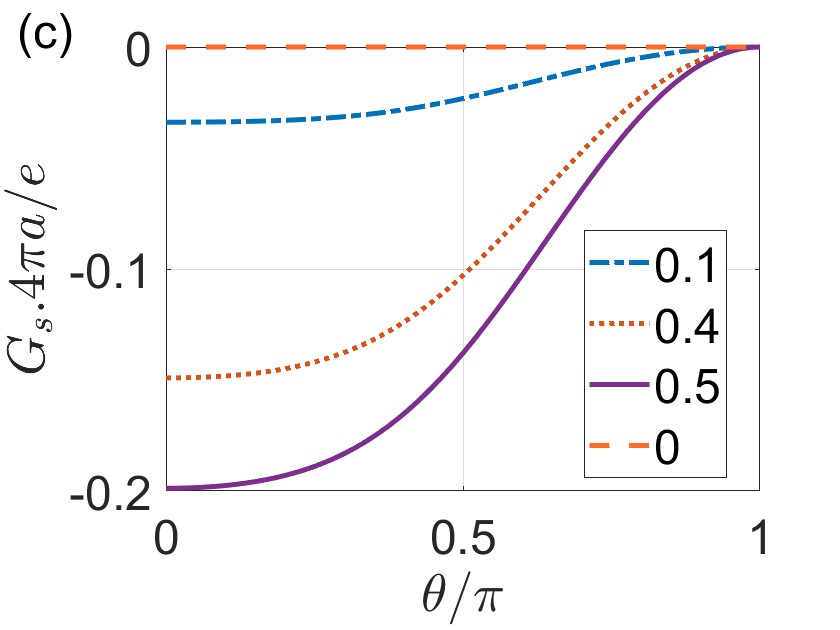}
\includegraphics[width=8cm]{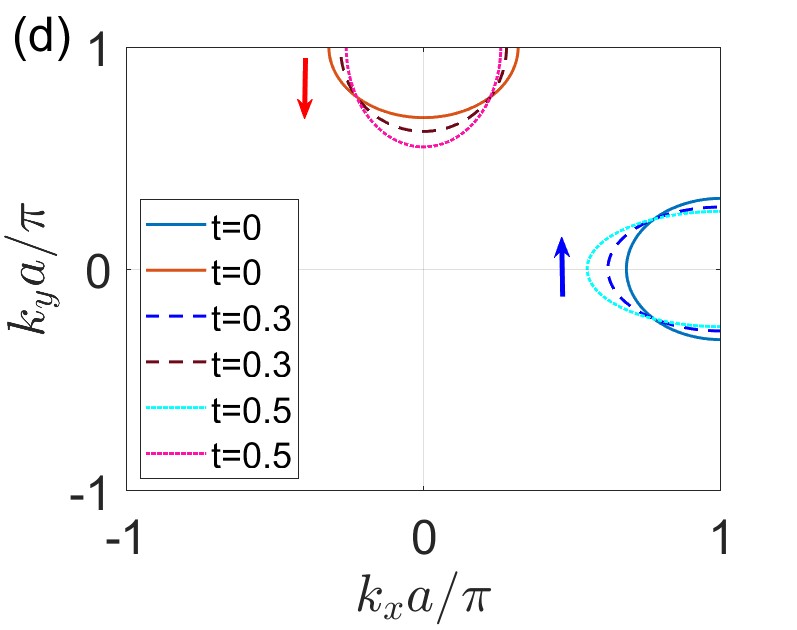}
\caption{(a) Schematic of the system. The curves indicate Fermi surface. The N\'eel vectors on the left AM and right AM differ by an angle $\th$. (b) Differential charge conductivity versus $\th$ and (c) spin conductivity versus $\th$ for different values of $t_J$ as indicated in the legend. Other parameters: $q_0=1/a$, $c=1$, $E=t_J$ (d) Fermi surface of up- and down-spin electrons for different values of $t$. }\label{fig:strong}
\end{figure}
Mn$_5$Si$_3$ is a candidate material for AM in the strong regime with $t_J=2t=~150meV$~\cite{Helena21}. 
Figure~\ref{fig:strong}(b) shows the variation of charge conductivity with the spin polarization angle $\theta$  for different values of $t$. In the strong phase, conductivity is dominated by a single spin channel, and electron transport occurs only when the transverse momentum matches across the two AM regions. As shown in Fig.~\ref{fig:strong}(a), an electron incident with a given spin is transmitted into the right AM with the same spin. So, there is no down-spin current for the up-spin electron incidence and vice versa. The conductivity is maximum at $\theta=0$ and decreases monotonically with increasing $\theta$, vanishing at $\theta=\pi$. This behavior arises because, for $\theta=0$, the spin orientations in both regions are collinear, enabling large transmission. With increasing $\theta$, the spin overlap is continuously reduced, thus suppressing the conductivity. At $\theta=\pi$, the spins are fully antiparallel, eliminating overlap and hence blocking transmission into the right AM, resulting in zero conductivity.

Figure~\ref{fig:strong}(c) illustrates the variation of spin conductivity with the spin polarization angle $\theta$ for different values of the hopping parameter $t$ on both sides of the interface. The spin conductivities corresponding to either side of the junction coincide for identical parameter values and decrease progressively to zero as $\theta$ approaches $\pi$. This behavior arises from the spin-dependent Fermi surface of the altermagnet. For $\theta=0$, the spin polarizations in the two AM regions are collinear, allowing efficient transmission of spin-polarized electrons and yielding maximum spin conductivity. As $\theta$ increases, the relative spin alignment between the two AMs is reduced, causing a mismatch between spin states and thereby suppressing the spin-resolved conductivities. Since there is no down-spin current for the up-spin incidence and no up-spin current for the down spin incidence, so for a particular $\th$ difference between the up- and down-currents remains the same on either sides of the junction. The complete overlap thus reflects the high symmetry of the AM Fermi surface, which ensures equivalent transport characteristics on both sides.

The spin conductivity, defined as the difference between up-spin and down-spin current, can take negative values which shows that contribution to the current due to down-spin electrons is higher than the up-spin electrons. This is explained by Fig. ~\ref{fig:strong}(d) where the Fermi surface for up- and down-spin is drawn for different values of $t$. It is clear from the above figure  that in the region $0<t<t_J$ as $t$ grows from 0 to $t_J$, the Fermi surface for up-spin electrons is shortened along $k_y$ occupying less number of transverse momentum states whereas the Fermi surface for down-spin electrons is elongated along $k_y$ occupying larger number of $k_y$ states resulting into higher down-spin conductivity for $t\neq0$. When $t=0$, the spin conductivity completely vanishes due to equal contribution of up-and down-spin electrons because at this value of $t$ Fermi surface for both the spins are exactly identical occupying same number of $k_y$ states.  Thus, the sign of the spin conductivity reflects the imbalance between spin-resolved transport channels.

\section{AM-NM-AM junction in strong phase} \label{sec:ANAS}
  \subsection{Details of the calculation}
   Now a NM of length $L$ is sandwiched between the two AMs having different N\'eel vectors characterized by $\chi$. The Hamiltonian in region to the left of $x=0$ is $H_S(\chi=0)$, whereas to the right of  $x=L$ is $H_S(\chi=\theta)$. In the region between $0<x<L$, the Hamiltonian for the NM is given below - 

\begin{align}
    H_{nm}&=-t_0\sigma_0a^2(\Do_x^2+\Do_y^2)-\mu~~~~~{ \rm for}~~(0<x<L)
    \label{eq:HN-weak}
\end{align}
Conservation of current probability density at the two interfaces $x=0~~ {\rm and}~~ x=L$ results in  boundary condition.
\bea 
\psi(0^-) &=& c\psi(0^+) \nn \\ 
c\big[(t_J\si_0-t\si_z)(a\Do_x-i\pi\si_{\ua})\psi\big]_{0^-} &=& \big[t_0\si_0(a\Do_x-aq_0)\psi\big]_{0^+} \nn \\
\psi(L^-) &=& c\psi(L^+) \nn \\ 
c\big[t_0(\si_0a\Do_x+aq_0)\psi\big]_{L^-} &=& \big[(t_J\si_0-t\si_{\theta})(a\Do_x \nn \\ && -i\pi\si_{\ua\theta}\psi\big]_{L^+} \nn \\
&& \label{eq:bc-strng_NM}\eea

A $\ua$-spin electron with energy $E$, incident from the left AM at an angle $\al$ relative to the $x$-axis, is described by a scattering wavefunction of the form $\psi(x)e^{ik_{y\ua} y}$, where 
\bea 
\psi(x) &=& (e^{ik_{x\ua}x}+r_{\ua\ua}e^{i(2\pi/a-k_{x\ua})x})\ket{\ua} +r_{\da\ua}e^{-ik_{x\da }x}\ket{\da}, \nn \\ 
&& {\rm for ~~} x<0, \nn \\ 
&=& \Big(A_R e^{iq_xx} + A_L e^{-iq_xx}\Big)\ket{\ua}+\Big(B_R e^{iq_xx} + \nn \\ &&  B_L e^{-iq_xx}\Big)\ket{\da} ~~~~ {\rm for~~} 0<x<L \nn \\
&=& t_{\ua\ua}e^{ik_{x\ua}x}\ket{\ua_{\th}} +t_{\da\ua}e^{ik_{x\da}x}\ket{\da_{\th}},\nn \\ && {\rm for~~} x>L 
\eea
where the expressions for $k_{x\ua}a,~k_{x\da}a~{\rm and}~ k_{y_\ua}$ are given by  equation \eqref{eq:wv_strng} and $q_xa=\sq{(E+\mu)/t_0-k_{y\ua}^2}$. The scattering coefficients $A_R,~B_R,~A_L,B_L,~r_{\si'\si}$ and $t_{\si'\si}$ can be found using the boundary conditions in Eq.~\eqref{eq:bc-strng_NM}.\\

The charge- and spin-  current densities on two sides of the junction due to this wavefunction are given by Eq. \eqref{eq:Jup-strng}, where $J^{s-}_{\da}(\al)$ and $J^{s+}_{\da}(\al)$ are the spin current for the left and right AM respectively.

Similarly a $\da$-spin electron with energy $E$, when incident from the left AM at an angle $\al$ with respect to the $x$-axis, is described by a scattering wavefunction of the form $\psi(x)e^{ik_{y\da} y}$, where
\bea 
\psi(x) &=& (e^{ik_{x\da}x}+r_{\da\da}e^{-ik_{x\da}x})\ket{\da} +r_{\ua\da} e^{i(2\pi/a-k_{x\ua})x}\ket{\ua}, \nn \\ 
&& {\rm for ~~} x<0, \nn \\ 
&=& \Big(A_R e^{iq_xx} + A_L e^{-iq_xx}\Big)\ket{\ua}+\Big(B_R e^{iq_xx} + \nn \\ &&  B_L e^{-iq_xx}\Big)\ket{\da} ~~~~ {\rm for~~} 0<x<L \nn \\
&=& t_{\ua\ua}e^{ik_{x\ua}x}\ket{\ua_{\th}} +t_{\da\ua}e^{ik_{x\da}x}\ket{\da_{\th}},\nn \\ && {\rm for~~} x>L .
\eea
where the expressions for $k_{x\da}a,~k_{x\ua}a~{\rm and}~ k_{y_\da}$ are given by  equation \eqref{eq:wv_weak} and $q_xa=\sq{(E+\mu)/t_0-k_{y\da}^2}$.  Using the boundary conditions in Eq.~\eqref{eq:bc-strng_NM} the scattering coefficients  $A_R,~B_R,~A_L,B_L,~r_{\si'\si}$ and $t_{\si'\si}$ are calculated. \\

The charge- and spin-  current densities on two sides of the junction due to this wavefunction are given by Eq. \eqref{eq:Jdn-strng}, where $J^{s-}_{\da}(\al)$ and $J^{s+}_{\da}(\al)$ are the spin currents for the left and right AM respectively. The differential charge and spin conductivities in this system are calculated using  Eq.\eqref{eq:cond_wk}.

\subsection{Results}
\begin{figure}[htb]
\includegraphics[width=8cm]{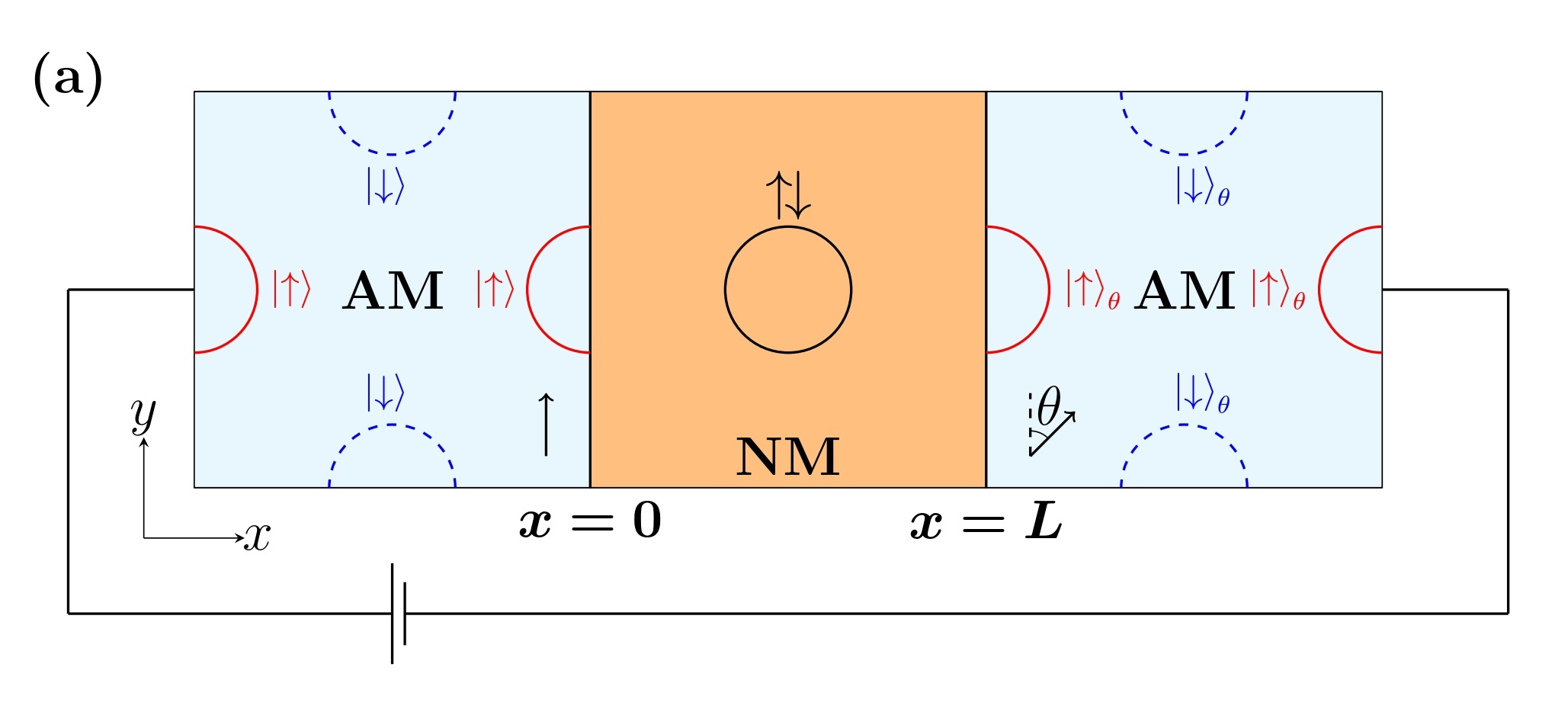}
\includegraphics[width=4cm,height=3.5cm]{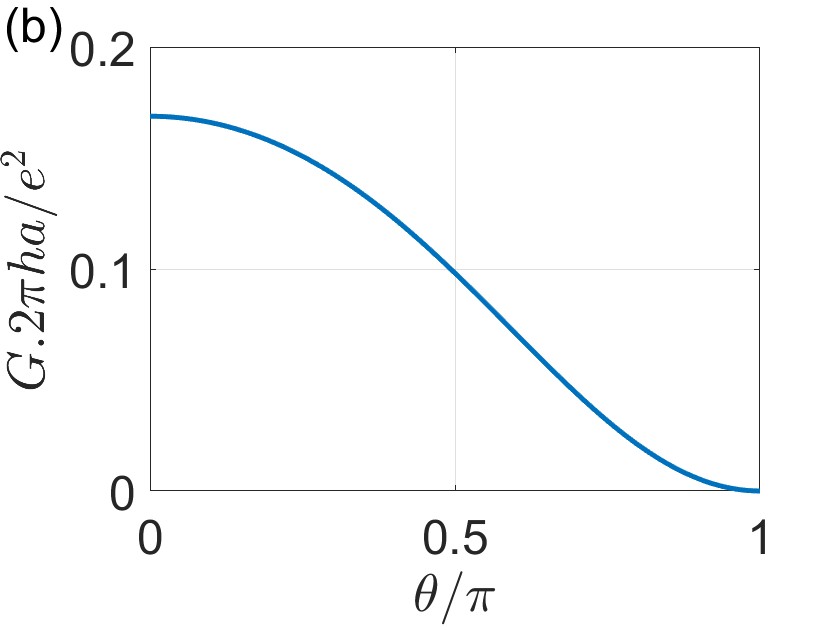}
\includegraphics[width=4cm,height=3.5cm]{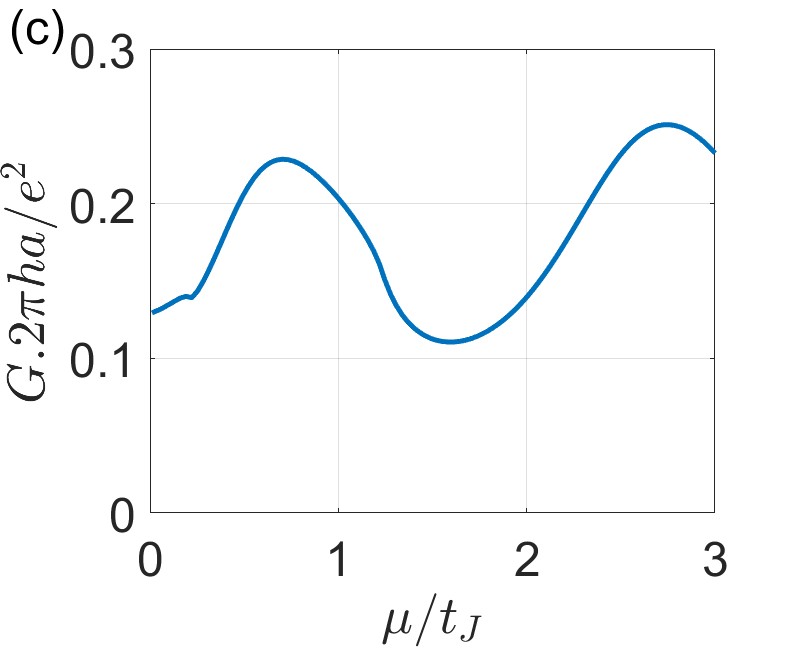}
\includegraphics[width=4cm,height=3.5cm]{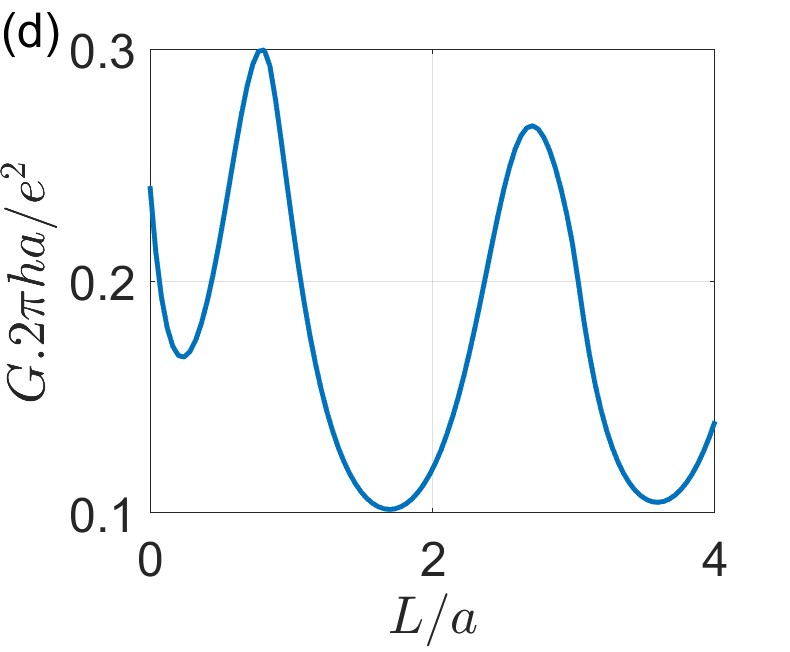}
\caption {(a) Schematic of the system. The curves show Fermi surfaces in each region. The N\'eel vectors on the left AM and right AM differ by an angle $\th$.  Total charge conductivity (b) versus $\th$ keeping $L=3a$ and $\mu=t_J$ (c) versus $\mu$ keeping $L=4a$ and $\th=0$, and (d) versus $L$ keeping $\mu=2t_J$ and $\th=0$ for $\ua$ and $\da$ spin electrons indicated in the legend. Other parameters: $q_0=1/a$, $c=1$, $t=0.1t_J$, $E=t_J$.}\label{fig:strong_NM}
\end{figure}
Figure~\ref{fig:strong_NM}(b) shows total charge conductivity  with respect to spin polarization angle $\theta$.  In particular, the up-spin conductivity is maximum at $\theta = 0$ and decreases gradually to zero as $\theta$ approaches $\pi$. The spin-dependent band structure and transverse momentum matching across the junction explains this behaviour. At $\theta = 0$, the spin polarization in both the AMs is aligned along the same direction, allowing large transmission of up-spin electrons. As $\theta$ increases, the spin alignment between the two AM regions deviates, introducing a spin mismatch that reduces the probability of transmission of the up-spin electrons. At $\theta = \pi$, the magnetizations are fully anti-aligned, and the right AM supports only down-spin states, thereby completely blocking the transmission of up-spin electrons and resulting in zero conductivity. On the other hand conductivity due to the down-spin electrons is completely zero irrespective of $\theta$ because of mismatching of transverse momentum at the junction of left AM and NM. So the down-spin transmission suppressed exponentially. \\

Figure~\ref{fig:strong_NM}(c) shows variation total charge conductivity with respect to the chemical potential $\mu$. Here also, we observe that the conductivity due to down-spin incident electrons remains zero for all values of $\mu$, reflecting the same underlying spin-dependent transport mechanism described earlier. In contrast, up-spin incident electrons exhibit finite conductivity in the right AM region, which shows an oscillatory dependence on $\mu$ due to FPI effects in the normal metal.  As  $\mu$ varies, the corresponding change in the Fermi wavevector alters the phase accumulation of the electron wavefunction due to back and forth reflection within the NM. This results in either constructive or destructive interference, thereby causing oscillations in the transmission probability and hence in the conductivity. The FPI condition, $\De q=\pi/L$,  gives 0.72/a in comparison to 0.68/a that is found  in Fig.~\ref{fig:strong_NM}(c) at two different chemical potential  $\mu_1$ and $\mu_2$ measured at the peaks calculated by $\De q=\sqrt{(E+\mu_2)/t_0}-\sq{(E+\mu_1)/t_0}$.

Figure~\ref{fig:strong_NM}(d) shows the variation of total charge conductivity with respect to the length $(L)$ of the NM. The oscillatory behavior of the conductivity wavevector $L$ arises due to FPI caused by multiple reflections of electron wavefunctions within the NM region. Similar to the above cases, the FPI condition, $\De L=\pi/q_x$, at normal incidence gives  2.22a, whereas the numerically obtained value observed in the results from the Fig.~\ref{fig:strong_NM}(d) is 2a. This discrepancy arises because the Fabry–P\'erot interference condition applied above assumes normal incidence, but actually the electrons also propagate at oblique angles, each with its own distinct interference condition. Most of the contribution to the conductivity is due to the up-spin transmitted electrons. Down-spin electrons contribute only at smaller lengths and decay exponentially inside the NM as evanescent modes due to transverse momentum mismatch at the interface. 

\section{Discussion}~\label{sec:disc}
Transport across AM-NM interfaces has been shown to generate spin currents~\cite{das2023}. Moreover, spin-resolved transport and spin-valve effects have been demonstrated in NM-AM-NM junctions~\cite{fu2025am}. In contrast, the present work focuses on two different junction geometries: (i) AM-AM and (ii) AM-NM-AM, and systematically explores how charge and spin transport depend on the altermagnetism and the relative orientation of the N\'eel vectors.

TMR defined as the ratio of the difference between the conductivities in parallel and antiparallel alignment of the two AMs to the smaller of the two (see ref.~\cite{Sun25} for this definition) shoots up to infinity for the strong phase since the conductivity  in the antiparallel configuration is zero and that in the parallel configuration is finite. We find that TMR in the weak phase turns out to be of the order of $4-12\%$ for AM-NM-AM junctions. 

Realistic heterostructures inevitably contain defects, disorder, and edge roughness, which can influence transport properties. In the present work, our analysis is based on a continuum model in the ballistic regime, where translational invariance along the transverse ($\hat y$) direction is assumed. Introducing impurities or rough edges breaks this invariance, making such effects beyond the scope of the current framework. A lattice model would be more appropriate for incorporating disorder and edge roughness in a controlled manner.

\section{Summary and conclusion} \label{sec:sum}

We have investigated electron transport across junctions involving altermagnets  by analyzing two distinct regimes - the strong and weak phases. For each case, we calculate the charge and spin conductivities as functions of the angle between the N\'eel vectors  $\theta$.
In the strong AM regime, the total charge conductivity gradually decreases and eventually vanishes as  $\theta$ approaches $\pi$. In contrast, for the weak AM case, the charge conductivity remains finite even at $\theta = \pi$, reflecting the partial spin polarization characteristic of the weak phase. Similarly, the spin conductivity on the left and right AM electrodes are identical in the strong AM regime, signifying symmetric spin transport, whereas in the weak AM phase this symmetry is lost except when $\theta = 0$.

To further explore the transport behavior, we introduced a normal metal between the two AM layers and studied the variation of charge conductivity versus chemical potential and the NM length for both phases. The chemical potential of the central NM can be changed by application of an external gate voltage. The conductivity exhibits clear Fabry–P\'erot type oscillations originating from quantum interference between multiple reflections at the AM/NM interfaces. The oscillation frequency is noticeably higher in the weak AM case and lower in the strong AM, consistent with the difference in the effective wavevectors and spin-dependent potentials of the two regimes.
Moreover, in the strong AM phase, the transport is almost completely dominated by up-spin electrons, demonstrating strong spin selectivity, while in the weak AM phase, both spin channels contribute significantly to the overall conductivity. The charge and spin conductivites of the system are tunable in altermagnetic materials, by varying the angle between N\'eel vectors of two AMs.

Recent theoretical proposals have outlined mechanisms to control the N\'eel vector in altermagnets~\cite{han25,zhang25}. These advances indicate that the phenomena we predict may soon be experimentally testable and could play a role in guiding the development of future spintronic device architectures.
\acknowledgements
SG, SD and AS thank  Science and Engineering Research Board (now Anusandhan National Research Foundation) Core Research grant (CRG/2022/004311) for financial support.

\bibliography{ref_almag}
\end{document}